\documentstyle[lanl,pslatex,pslatex,rawfonts]{article}  

\frompage{000} \topage{000}                                              
\def\rightmark{Dissipative Relativistic Fluid Dynamics for Nuclear Collisions} 
\def\leftmark{Azwinndini Muronga}

\newcommand{\nc}{\newcommand}
\nc{\bce}{\begin{center}}
\nc{\ece}{\end{center}}
\newcommand{\eps}{{\varepsilon}}
\newcommand{\btu}{{\bigtriangleup}}
\newcommand{\btd}{{\bigtriangledown}}
\title{{\small \rightline{NUC-MINN-01/3-T}}\vspace{.2cm}
Dissipative Relativistic Fluid Dynamics for Nuclear Collisions} 
\authors{
{Azwinndini Muronga %
 }\\[2.812mm]
{\normalsize
\hspace*{-0pt}School of Physics and Astronomy, University of Minnesota, \\ 
116 Church Street S.E., Minneapolis, Minnesota 55455, USA\\[0.2ex] 
Email address: amuronga@physics.spa.umn.edu
}}
 
\abstract{In the 
context of the M\"uller-Israel-Stewart second-order theory for dissipative
fluids due to Grad, we analyze the effects of thermal conduction and viscosity
in heavy ion collisions. We contrast the results to those of the first-order
theory due to Eckart and to Landau and Lifshitz and to those of perfect (ideal)
fluid due to Euler. We study the energy density and entropy density evolution of a
pion gas produced in the heavy ion collisions. The truncated version of the
second-order theory is used to find the dissipative quantities.}

 
\begin{document}
 
\maketitle

\section{Introduction}\label{intro}
Heavy ion collisions such as those at RHIC provide a basic tool to
study the properties of hot and dense matter produced at high energies.
Of particular importance is knowledge of the space-time evolution of the
matter. This is important when one wants to study the transition from hadron to
quark and gluon degrees of freedom, a state of matter known as QGP (quark-gluon
plasma), as predicted by QCD \cite{azwi:qgpliterature}. 
One way to study the dynamics of the produced matter is by using fluid dynamics
\cite{azwi:stocker1986,azwi:kajantie1986}. To probe the non-equilibrium properties of the matter we
need to include dissipative effects in the fluid dynamic modeling of heavy
ion reactions.
It is known even in non-relativistic dissipative fluids that dissipation will
have an effect on the observables \cite{azwi:kapusta}.

The standard theory of dissipative fluid dynamics developed by Eckart
\cite{azwi:eckart1940} and by Landau and Lifshitz \cite{azwi:landau1959}
exhibits certain undesirable effects. The resulting transport equations of dissipative fluxes
lead to parabolic equations for heat conduction and shear diffusion. Thus it 
predicts an infinite speed of propagation
for thermal and viscous signals. Many
applications of dissipative fluid dynamics in relativistic nuclear collisions
have used the standard theory \cite{azwi:naviergang}. 
Because of such undesirable effects, it is then
necessary to apply another thermodynamic theory of irreversible processes
that does not present this anomalous behavior. Causal theory of
dissipative fluids developed by M\"uller \cite{azwi:muller1967} and by
Israel and Stewart \cite{azwi:israelstewart}  due to Grad \cite{azwi:grad1949} 
were developed to remedy some of
these undesirable features. The resulting equations of causal theory are hyperbolic
in structure and lead to causal propagation. Thus causal theory seems
to be a good candidate to use instead of the standard theory. 
In the causal-type theories the space of
thermodynamic variables is extended to include dissipative flows. These 
dissipative flows (heat flows and viscous pressures) are considered as
independent variables. The entropy four-flow then depends not only on the
primary variables (number density, energy density, and pressure) but on these
dissipative flows as well and its production
is semi-positive definite. At equilibrium, the entropy is maximum.

\section{Basic Features of Relativistic Dissipative Fluid Dynamics}\label{basics}
 
The basic formulation of relativistic hydrodynamics can be found in literature 
\cite{azwi:landau1959,azwi:weinberg,azwi:thorne,azwi:laszlobook,azwi:degroot,azwi:rischke}. 
We will use the natural units: $\hbar = c = k_{\rm B} = 1$.
The metric tensor we use is $g^{\mu \nu} = \mbox{diag} (+,-,-,-)$.
We consider a simple fluid and no electromagnetic
fields. This fluid is characterized by,
\begin{eqnarray}
N_A^\mu(x) & & \makebox[0.3in][l]{}\mbox{(particle 4-current)}\enspace,\\
T^{\mu\nu}(x) & & \makebox[0.3in][l]{}\mbox{(energy-momentum tensor)} \enspace,\\
S^\mu(x) & & \makebox[0.3in][l]{}\mbox{(entropy 4-current)} \enspace,
\end{eqnarray}
where $A = 1,...,n$ for the $n$ conserved net charge currents, such as electric
charge, baryon number, and strangeness.
$N^\mu$ and $T^{\mu\nu}$ represent conserved quantities:
\begin{eqnarray}
\partial_\mu N_A^\mu &\equiv& 0 \enspace, \\
\partial_\mu T^{\mu\nu} &\equiv& 0\enspace.
\end{eqnarray}
The above equations are the local conservation of net charge and
energy-momentum. They are the equations of motion of a relativistic fluid. There
are $4+n$ equations and $10+4n$ independent unknown functions. 
The second law of thermodynamics requires
\begin{equation}
\partial_\mu S^\mu \geq 0 \enspace,
\end{equation}
and it forms the basis for the development of the extended irreversible
thermodynamics.

\subsection{Tensor Decomposition}\label{details}

We now perform a tensor decomposition of $N_A^\mu,\,T^{\mu\nu},$ and $S^\mu$ with
respect to an arbitrary, time-like, normalized 4-vector $u^\mu$, $u^\mu u_\mu
=1.$ The projection onto the 3-space orthogonal to $u^\mu$ is denoted by
\begin{equation}
\btu^{\mu \nu} \equiv g^{\mu \nu} - u^\mu u^\nu =
\btu^{\nu \mu}\, ,\,\,
\btu^{\mu \nu} u_\nu = 0 \, , \,\, \btu^{\mu \alpha} 
\btu_{\alpha}^\nu = \btu^{\mu \nu}\,,\,\,\,
\btu^\nu_\nu = 3  \enspace.
\end{equation}
The tensor decomposition reads:
\begin{eqnarray}
N^\mu_A & = & n_A \, u^\mu + V_A^\mu \enspace , \\
T^{\mu \nu} & = & \eps \, u^\mu u^\nu - p\, \btu^{\mu \nu}
+ 2\,W^{(\mu }\,u^{\nu)} + t^{\mu \nu} \enspace,\\
S^\mu &=& s u^\mu + \Phi^\mu \enspace,
\end{eqnarray}
where we have defined
\begin{eqnarray}
W^\mu &=&  q^\mu + h\, V^\mu \enspace,\\
t^{\mu\nu} &=& \pi^{\mu\nu} - \Pi\, \btu^{\mu \nu} \enspace,
 \end{eqnarray} 
and the parenthesis notation is defined by
\begin{equation}
A^{(\mu\nu)} \equiv \frac{1}{2} \left(A^{\mu\nu} + A^{\nu\mu}\right) \enspace.
\end{equation}
In this presentation $h$ is the enthalpy per particle defined by
\begin{equation}
h = \frac{(\eps + p)}{n} \enspace.
\end{equation} 
The dissipative fluxes satisfy the following orthogonality relations:
\begin{equation}
u_\mu V^\mu_A=0\,,\,\,u_\mu q^\mu=0\,,\,\,u_\mu\, W^\mu =0\,,\,\, u_\mu\,
t^{\mu\nu} =0 \,, \,\, \pi^\nu_\nu =0 \enspace.
\end{equation}

In the local rest frame (LRF) where $u^\mu = (1,\vec{0})$, the quantities appearing
in the decomposed tensors have the following meanings:
\begin{eqnarray}
n_A & \equiv & \makebox[1.in][l]{$u_\mu N_A^\mu$} \mbox{net density of charge of type
$A$}\enspace,\\
V_A^\mu &\equiv& \makebox[1.in][l]{$\btu^\mu_\nu N_A^\nu$} \mbox{net flow of charge of type
$A$}\enspace,\\
\eps &\equiv& \makebox[1.in][l]{$u_\mu T^{\mu\nu} u_\nu$} \mbox{energy density}\enspace,\\
p + \Pi &\equiv& \makebox[1.in][l]{$\displaystyle -\frac{1}{3} \btu_{\mu\nu} T^{\mu\nu}$ } \mbox{pressure}\enspace,\\
q^\mu &\equiv& \makebox[1.in][l]{$u_\nu T^{\nu\lambda} \btu^\mu_\lambda $} \mbox{heat flow}\enspace,\\ 
\pi^{\mu\nu} &\equiv& \makebox[1.in][l]{$T^{\langle\mu\nu\rangle}$ }\mbox{stress
tensor}\enspace,\\
s &\equiv& \makebox[1.in][l]{$u_\mu\, S^\mu $}\mbox{entropy density}
\enspace,\\
\Phi^\mu &\equiv& \makebox[1.in][l]{$\btu^\mu_\nu S^\nu $} \mbox{entropy flux} \enspace.
\end{eqnarray}
The angular bracket notation is defined by
\begin{equation}
A^{<\mu\nu>} \equiv \left[\frac{1}{2}\left(\btu^\mu_\sigma
\btu^\nu_\tau
+\btu^\mu_\tau \btu^\nu_\sigma\right)
-\frac{1}{3}\btu^{\mu\nu}\btu_{\sigma\tau}\right]
A^{\sigma\tau} \enspace.
\end{equation}
The space-time derivative decomposes into
\begin{equation}
\partial^\mu = u^\mu D + \btd^\mu \,, ~~~~~u^\mu \btd_\mu
=0 \enspace,
\end{equation}
where 
\begin{eqnarray}
D &\equiv& \makebox[1.in][l]{$u^\mu\partial_\mu $} \mbox{convective time
derivative}\enspace,\\
\btd^\mu &\equiv& \makebox[1.in][l]{$\btu^{\mu\nu}\partial_\nu $} \mbox{gradient
operator}\enspace.
\end{eqnarray}
In LRF the two operators attain the given meanings.
In this rest frame the projector becomes
\begin{equation}
\btu_{\mu\nu} =\btu^{\mu\nu} =
\mbox{diag}(0,-1,-1,-1),~~~~~~~~~~\btu^\mu_\nu = \mbox{diag}(0,1,1,1) \enspace,
\end{equation}
and the heat four flow has spatial components only ($q^\mu =(0,\vec{q})$).

\section{First Set of Equations: The Conservation Laws}\label{Laws}

From now on we will consider one type of charge, namely the net baryon number.
With the help of the orthogonality properties given in the previous section we obtain the
conservation laws.
The equation of continuity (baryon conservation), $\partial_\mu N^\mu \equiv
0$, the equation of motion (momentum conservation), $\btu^\mu_\nu\partial_\lambda
T^{\nu\lambda} \equiv 0$, and the equation of energy (energy conservation),
$u_\mu \partial_\nu T^{\mu\nu} \equiv 0$ are respectively given by
\begin{eqnarray}
D\,n &=& -n\,\btd_\mu \,u^\mu -\btd_\mu V^\mu + V_\mu D
\,u^\mu \enspace,\\
(\eps + p +\Pi) D\, u^\mu &=& \btd^\mu (p+\Pi) -
\btd_\nu \pi^{\mu\nu} + \pi^{\mu\nu} D \,u_\nu
-[\btu^\mu_\nu D\, W^\mu + 2 W^{(\mu} \btd_\nu u^{\nu)}]
\enspace,\\
D\, \eps &=& -(\eps + p + \Pi) \btd_\mu\, u^\mu +
\pi^{\mu\nu} \btd_\nu u_\mu -\btd_\mu W^\mu +2 W^\mu D
u_\mu \enspace. \label{energyequation}
\end{eqnarray}
There are $5$ conservation equations and $14$ unknown functions. We need
$9$ additional equations to close the system. So far $u^{\mu}$ is arbitrary. It can be chosen to be the
particle 4-velocity. This is known as the {\em Eckart frame} or {\em particle frame}. In this
frame $V^\mu \equiv 0$. 
Alternatively one can choose it to be the 4-velocity of the
energy flow. This is known as the {\em Landau and Lifshitz frame} or {\em energy frame}.
In this frame $W^\mu \equiv 0$.
Using the fundamental thermodynamic equation of Gibbs 
the energy balance equation can be written in the following convenient form:
\begin{equation}
T \partial_\mu S^\mu = \sigma_{\mu\nu}\pi^{\mu\nu} -\Pi \theta -\partial_\mu
q^\mu +q^\mu a_\mu \enspace, \label{entropyequation}
\end{equation}

where \begin{eqnarray}
        a_\alpha &\equiv & \makebox[2.in][l]{$u^\beta \partial_\beta u_\alpha$}  \mbox{(4-acceleration of
	the fluid)}\enspace,\nonumber\\
	\omega_{\alpha\beta} & \equiv & \makebox[2.in][l]{$ \displaystyle \btu_\alpha^\mu
	\btu_\beta^\nu\frac{1}{2}(\partial_\nu u_\mu - \partial_\mu u_\nu)$} 
	\mbox{(vorticity tensor)}\enspace,\nonumber\\
	\theta_{\alpha\beta} & \equiv & \makebox[2.in][l]{$ \displaystyle \btu_\alpha^\mu\btu_\beta^\nu
	\frac{1}{2}(\partial_\nu u_\mu + \partial_\mu u_\nu)$} \mbox{(expansion
	tensor)}\enspace,\\
	\theta & \equiv & \makebox[2.in][l]{$  \btu^{\alpha\beta}
	\theta_{\alpha\beta} = \partial_\alpha
	u^\alpha $} \mbox{(volume expansion) } \enspace,\nonumber\\
	\sigma_{\alpha\beta} &\equiv & \makebox[2.in][l]{$\displaystyle \theta_{\alpha\beta}
	-\frac{1}{3}\btu_{\alpha\beta}\theta $} \mbox{(shear
	tensor)}\enspace.\nonumber
\end{eqnarray}

\section{Second Set of Equations: The Transport Equations} \label{transport}

There are two approaches for finding the transport equations in addition to
conservation laws. The first one is the phenomenological approach and is based on the
second law of thermodynamics, that is, the principle of non-decreasing entropy. 
The standard theories of dissipative fluid dynamics
\cite{azwi:eckart1940,azwi:landau1959} assume that the entropy four flow is linear in dissipative
quantities. Hence they are also referred to as {\em first-order} theories.
The extended theories allow the inclusion of terms that are quadratic in the 
dissipative quantities, 
$\Pi\,,q^\mu$ and $\pi^{\mu\nu}$. They are referred to as {\em second-order} 
theories. 
Using conservation laws $u_\nu\partial_\mu T^{\mu\nu} = 0$ and $\partial_\mu
N^\mu =0$  we then obtain the expression for the entropy production
$\partial_\mu S^\mu$ from which 
we obtain the transport equations by requiring that the entropy production be
positive.
The second approach uses kinetic theory and is based on Boltzmann moment
equations. The results of transport equations to be presented here are obtained
by using a relativistic Grad's 14 moment approximation \cite{azwi:grad1949}.
However, one can also use the first Chapman-Enskog approximation
\cite{azwi:chapman} to find the transport
equations in the Eckart theory. 
Both the phenomenological and kinetic theory approaches require 
that the deviations from local
thermodynamic equilibrium be small, that is, $V^\mu,q^\mu,\pi^{\mu\nu}$ and $\Pi$
are small compared to $\eps,p,$ and $n$.

In the standard Eckart theory one obtains the following set of transport
equations for the {\em bulk viscous pressure}, the {\em heat flow} and the 
{\em shear viscous pressure} respectively
\begin{eqnarray}
\Pi &\equiv& -\zeta \btd_\mu u^\mu \enspace,\\
q^\mu &\equiv& \lambda T \left(\frac{\btd^\mu T}{T} - D u^\mu\right)
= -\lambda \, n \,T^2\,\btd^\mu \left(\frac{\mu}{T}\right)
\enspace,\\
\pi^{\mu\nu} &\equiv& 2 \eta \btd^{\langle\mu}u^{\nu\rangle}
\enspace,
\end{eqnarray}
where $\zeta(\sigma,\eps,n),\lambda(\sigma,\eps,n),$ and $\eta(\sigma,\eps,n)$ are the {\em bulk viscosity, thermal
conductivity} and {\em shear viscosity} coefficients. These transport
coefficients are required to be positive by the second law of thermodynamics
\begin{equation}
\partial_\mu S^\mu = \frac{\Pi^2}{\zeta T} - \frac{q^\mu q_\mu}{\lambda T^2} +
\frac{\pi^{\mu\nu}\pi_{\mu\nu}}{2 \eta T} \geq 0 \enspace.
\end{equation}
The resulting equations of motion are parabolic, 
unstable under perturbations and lead to
an acausal nature of propagation \cite{azwi:israelstewart,azwi:hiscock}. This is 
a paradox since in special relativity the speed of light is finite and
all maximum speeds should not be greater than this speed. This paradox was
first addressed by Cattaneo \cite{azwi:cattaneo}. In the above set of equations, 
if a thermodynamic force is suddenly switched off, then the corresponding flux 
instantaneously vanishes, indicating that a signal propagates through the
fluid at infinite speed, violating relativistic causality. Even in the
non-relativistic case, infinite speeds present a problem, since physically we
expect the signal speed to be limited by the maximum molecular speed. 
To avoid this paradox Cattaneo introduced {\em ad hoc} relaxation 
terms in the phenomenological equations. The resulting equations 
conform with causality and hyperbolicity requirements. The only problem was that a theory developed from first
principles was 
needed. It is from these arguments that the extended theory of M\"uller, Israel and
Stewart was developed. 

The resulting transport equations derived from extended theories differ
from those of Eckart-type theories: they contain relaxation terms. These
relaxation terms make the structure of the resulting equations hyperbolic and
thus will conform with causality requirements. 
Although it may not be reasonable in some situations, we shall assume here for
simplicity that there are no viscous/heat couplings (i.e. $\alpha_0 =\alpha_1 =0$
in \cite{azwi:israelstewart} ). We will also assume that there are no couplings 
of vorticity and acceleration to the heat and shear fluxes.
In these approximations we are left with a simple but still causal structure of
transport equations which has the Maxwell-Cattaneo \cite{azwi:cattaneo} form of
transport equations. 
The M\"uller-Israel-Stewart equations reduce, in the Eckart frame, to
\begin{eqnarray}
\tau_\Pi D \Pi +\Pi &\simeq& -\zeta \theta  \enspace,\\
\tau_q D q^\mu + q^\mu &\simeq& \lambda T 
\left(\frac{\btd^\mu T}{T} - D u^\mu \right)\enspace,\\
\tau_\pi D \pi^{\mu\nu} +\pi^{\mu\nu} &\simeq& 2\eta \sigma^{\mu\nu}\enspace ,
\end{eqnarray}
where
\begin{equation}
\tau_\Pi = \zeta \beta_0,~~~\tau_q = \lambda T \beta_1,~~~\tau_\pi = 2 \eta
\beta_2 \enspace.
\end{equation}
Here the $\beta_A(\eps,\,n)$ are the relaxation coefficients of the dissipative fluxes.
As before $\zeta,\,\lambda,\,\eta$ are the transport coefficients. 
They involve complicated collision integrals, and they also depend on the equation
of state. The $\tau_A$ are the relaxation times. They 
are sometimes taken to be the collision time ($\tau_{col} \approx 1/(n\,\sigma\,v)$,
with $\sigma$ being the cross section and $v$ the mean particle speed). In general they are 
different from the collision time. Here
$\tau_A$ will be taken to be the time for dissipative fluxes to relax to
their equilibrium values.
The relaxation terms in the extended theories make it possible for one to study the
evolution of the dissipative quantities.
The transport equations are coupled to the evolution equations for number
density, energy density and momentum. The evolution equations together with the
transport equations form a quasi-linear, symmetric and hyperbolic system of 14 
first order partial differential
equations. The system of equations is found to fulfill the requirements of causality and 
hyperbolicity \cite{azwi:israelstewart,azwi:hiscock}.
One also needs to investigate carefully the conditions under
which the truncated equations are reasonable.

\section{The equation of state and transport coefficients}\label{eos/transp}

\begin{figure}[htb]
\begin{center}
\begin{minipage}[b]{6.1cm}
           \epsfxsize 6.cm      \epsfbox{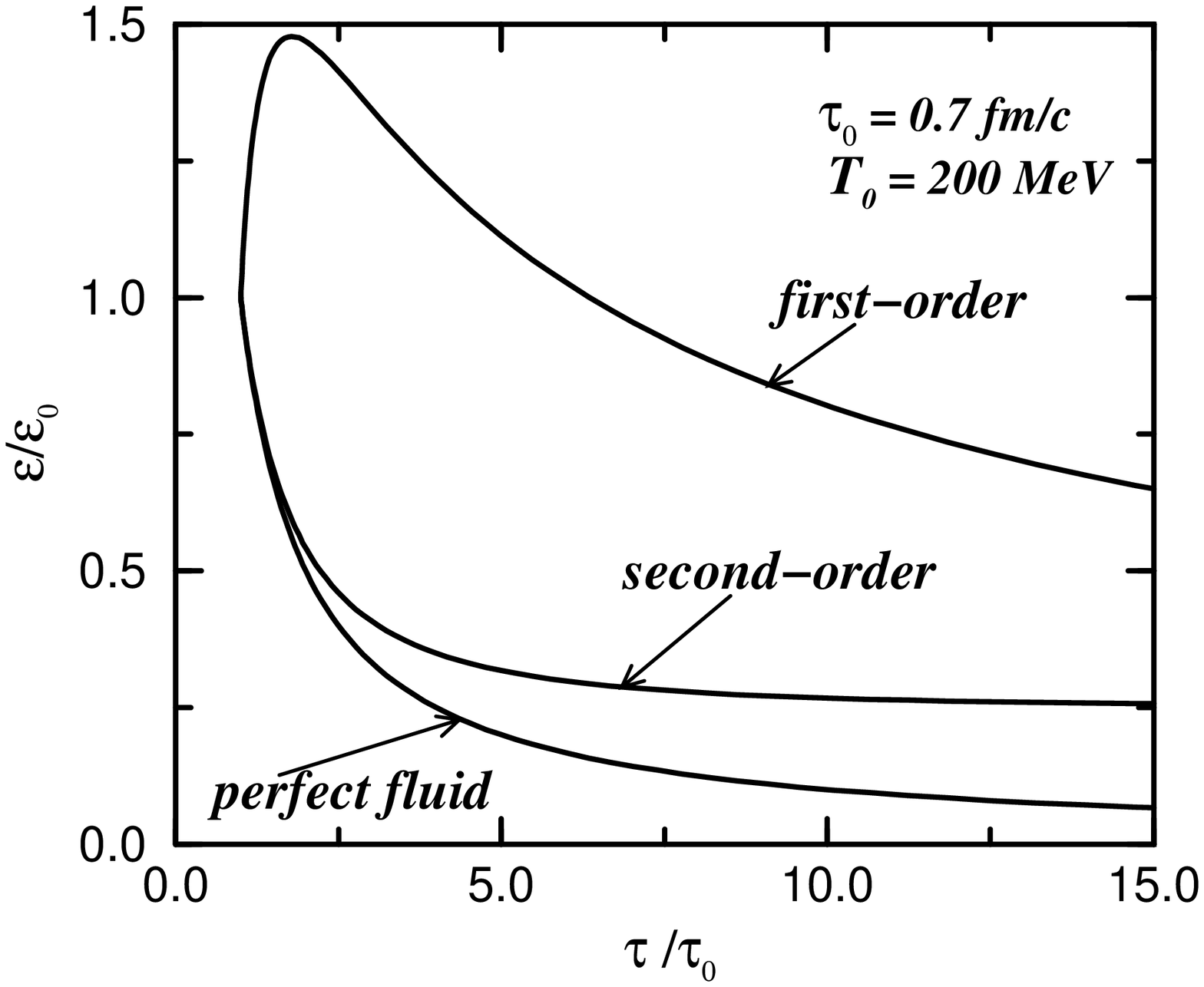}
\caption[]{The proper time evolution of energy density with initial conditions:
$\tau_0 = 0.7$ fm/c and $T_0= 200$ MeV.}
\label{azwi:fig1}
\end{minipage}
	   \hspace{.2cm} 
\begin{minipage}[b]{6.1cm}
           \epsfxsize 5.6cm      \epsfbox{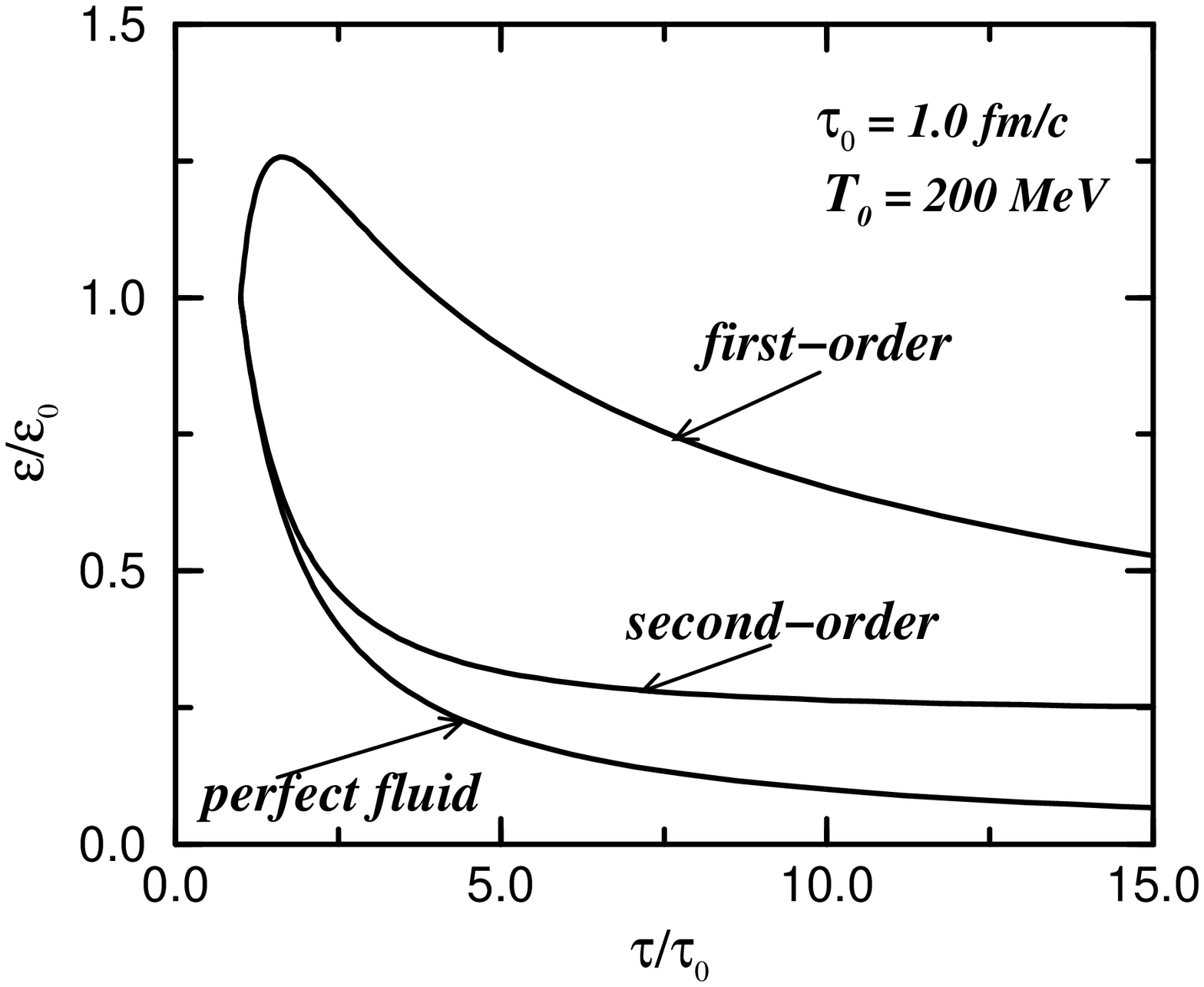}
\caption[]{The proper time evolution of energy density with initial conditions:
$\tau_0 = 1.0$ fm/c and $T_0= 200$ MeV.}
\label{azwi:fig2}
\end{minipage}
\end{center}
\end{figure}

\begin{figure}[htb]
\begin{center}
\begin{minipage}[b]{6.1cm}
           \epsfxsize 6.cm      \epsfbox{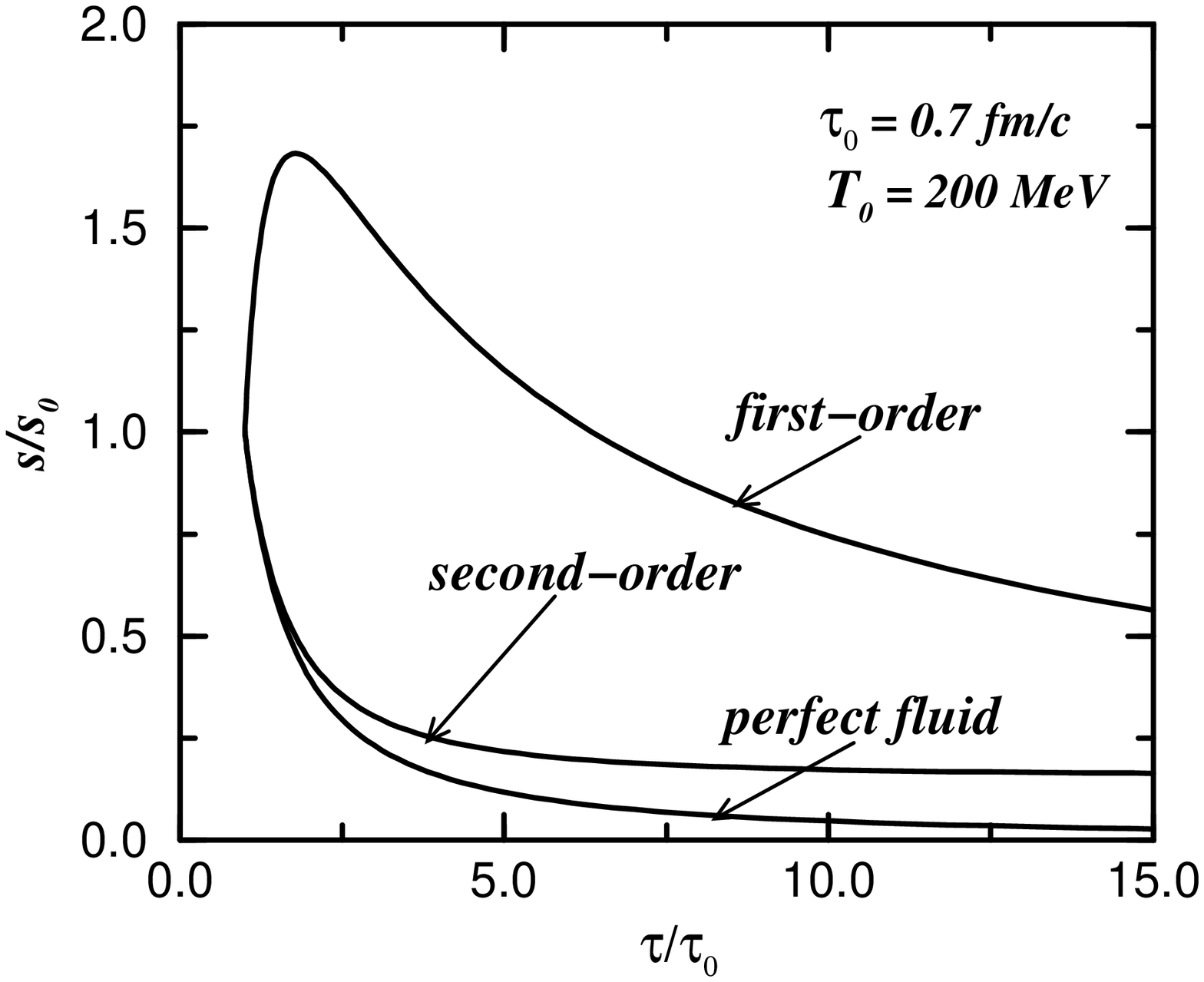}
\caption[]{The proper time evolution of entropy density with initial conditions:
$\tau_0 = 0.7$ fm/c and $T_0= 200$ MeV.}
\label{azwi:fig3}
\end{minipage}
	   \hspace{.2cm} 
\begin{minipage}[b]{6.1cm}
           \epsfxsize 5.6cm      \epsfbox{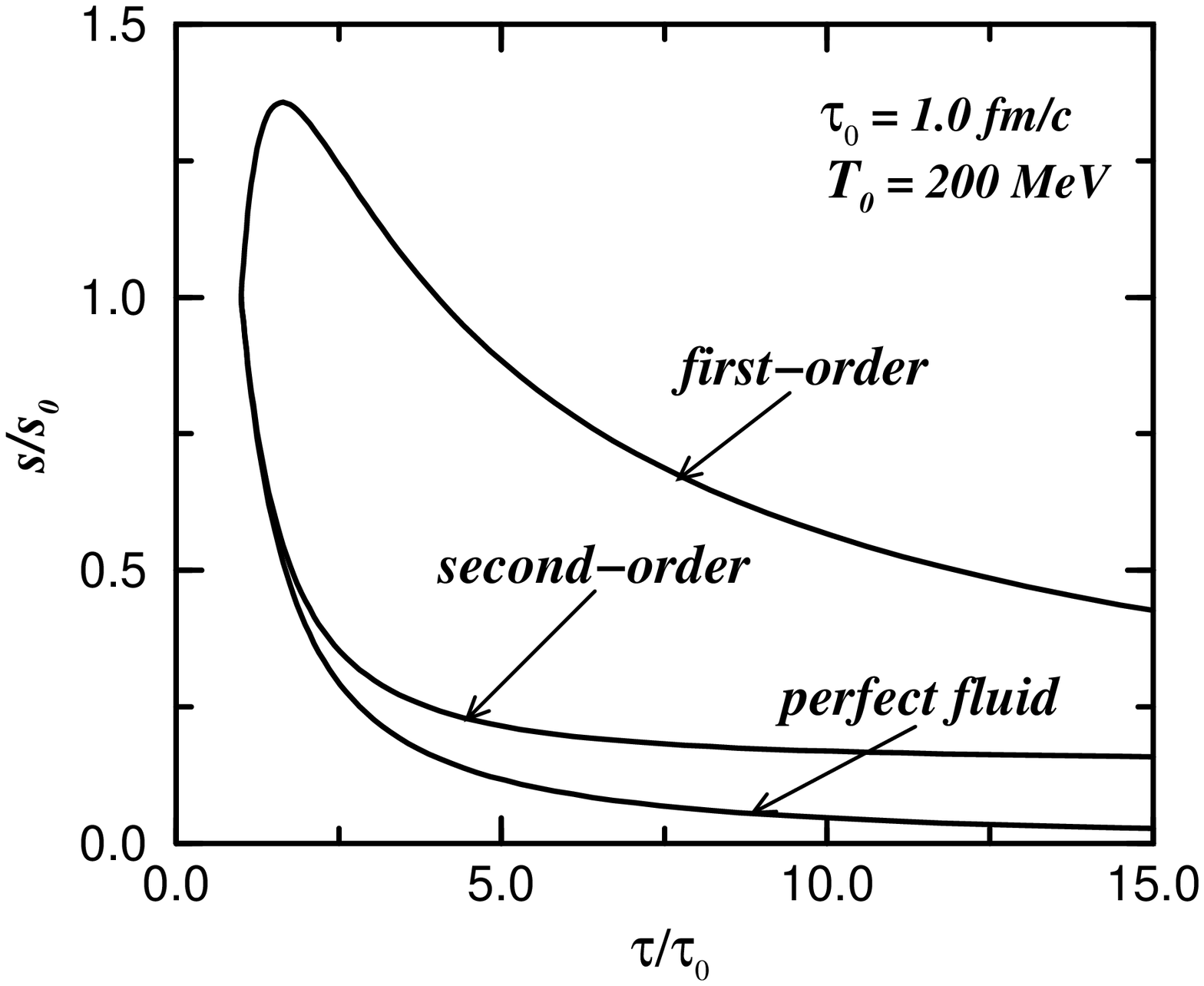}
\caption[]{The proper time evolution of entropy density with initial conditions:
$\tau_0 = 1.0$ fm/c and $T_0= 200$ MeV.}
\label{azwi:fig4}
\end{minipage}
\end{center}
\end{figure}

In this presentation we study the energy density and entropy density
evolution in the 1+1 Bjorken hydrodynamic limit \cite{azwi:bjorken}. 
We therefore consider equations (\ref{energyequation}) and
(\ref{entropyequation}). 
The equation of state is that of a massless pion gas. 
Thus the pressure is given by $p=a\,T^4$ with $a = g_h \pi^2/90$ 
where $g_h = 3$ is the number of degrees of freedom, the energy density 
and entropy density 
are given by $\eps= 3\,a\,T^4$ and $s=4\,a\,T^3$ respectively.
From the transport equations, the bulk viscous equation does not contribute for
massless particles, ($\zeta\longrightarrow 0$) \cite{azwi:weinberg}. For the 1+1
dimensional Bjorken-type hydrodynamics the heat term in the energy equation will
not contribute. Thus we need only the shear viscous pressure for this
presentation. 
The energy density evolution equation (\ref{energyequation}) becomes
\begin{equation}
\frac{d \,\eps}{d\,\tau} = -\frac{(\eps + p)}{\tau}
+\frac{\Phi}{\tau}
\end{equation}
where
\begin{eqnarray}
\Phi &\equiv& \makebox[1.in][l]{$0$} \mbox{(perfect fluid)} \enspace,\\
\Phi &=& \makebox[1.in][l]{$\displaystyle \frac{4}{3} \eta/\tau$} \mbox{(standard theory)}
\enspace,\\
\tau_\pi \frac{d\,\Phi}{d\,\tau} &=&  
\makebox[1.in][l]{$\displaystyle -\Phi + \frac{4}{3}\eta/\tau$} \mbox{(extended theory)} \enspace,
\end{eqnarray}
where $\eta = b\,T^{-1}$ \cite{azwi:prakash} and 
$(b = \pi/8 \,f_\pi^4)$ where $f_\pi$ = 93 MeV is the pion decay constant. 
For massless particles $\beta_2=3/(4\,p)$, and this is used
in the expression for $\tau_\pi$. 
The energy equation can be solved analytically for the
perfect fluid and first order (provided $\eta$ is constant) cases. But since we
want $\eta$ to depend on temperature or time one can then solve the equations
numerically or first find the temperature evolution as done in
\cite{azwi:azwi1}. 
In the case of the second-order theory we solve the equations numerically. The
proper time evolution of energy density is given by
\begin{eqnarray}
\eps(\tau) &=& \eps(\tau_0) \left[\frac{\tau}{\tau_0}\right]^{-4/3}
~~~~~~~~~~~~~~~~~~~~~~~~\mbox{(perfect fluid)} \enspace,\\
\eps(\tau) &=& \left\{\eps(\tau_0) -4\eta/\tau_0\right\}\left[\frac{\tau}{\tau_0}\right]^{-4/3}
+4\eta/\tau ~~~~~~~\mbox{(first order)} \enspace.
\end{eqnarray}
Here $\tau_0$ represents the instant at which the expansion starts.
In Figs. \ref{azwi:fig1} through \ref{azwi:fig4} we show the $\tau$ dependence of energy density
$\eps$ and entropy density $s$ for the three different cases: a perfect fluid, a first-order theory
of dissipative fluids 
and a  
second-order theory of dissipative fluids. In Figs. \ref{azwi:fig1} and
\ref{azwi:fig2} we show the dependence of the time evolution of energy density on
the initial time $\tau_0$. In Figs. \ref{azwi:fig3} and \ref{azwi:fig4} we show
the dependence of time evolution of entropy density on the initial time
$\tau_0$. In both cases we take the initial initial temperature to be 
200 MeV.
 
First we note that the presence of dissipation makes difference. In the earlier times 
the first-order (parabolic) theory predicts a peak in the energy density and
entropy while the perfect fluid and second-order (both hyperbolic) predict a 
monotonic decrease. By looking at the dependence of $\eps$ 
and
$s$ on $\tau_0$, we see that it is in the early times that the difference is most
pronounced. In the later times there is an indication that the 
two dissipative theories might converge. Therefore knowledge of different
time scales is crucial to determine when to apply which theory.
In heavy ion reactions, where the dynamics is happening in very short times, we need to use
the hyperbolic theories of relativistic dissipative fluid dynamics.

The presence of dissipation in heavy ion reactions will have profound effects on
the space-time evolution of the system. 
The freeze-out will be delayed. Temperature and energy density decrease slower. 
Enhancement of entropy production will increase the
production of final multiplicity since the two can be related. Since the system
takes longer to cool this will lead to an enhancement in the production of
thermal signals (dileptons and photons).

\section{Conclusions}

In the early stages of collisions, non-equilibrium effects play a 
dominant role. Thus, non-ideal fluid dynamics must be used to accurately describe the evolution of
the system. Because of the undesirable features of parabolic
theories, it is better to use the hyperbolic theories. 
Unlike in the first-order theories, where the transport equations are just the
algebraic relations between the dissipative fluxes and the thermodynamic forces, 
in the second-order theories the transport equations describe the evolution of the
dissipative fluxes from an arbitrary initial sate to a final steady-state. The
presence of the relaxation terms in second-order theories makes the structure of 
the resulting transport equations hyperbolic and thus have well-posed initial value
problems. 
The first challenge faced by the second-order theories is the increase in the space of
thermodynamic variables which brings new coefficients, in addition to transport coefficients, 
in the theory of non-ideal fluid dynamics. However, these new coefficients are
determined by the equation of state. Like the primary transport coefficients which
are constrained by the requirement of second law of thermodynamics, they are
constrained by the requirements of hyperbolicity and hence causality.
Thus, in principle, one still needs to know about the
transport coefficients and the equation of state in order to solve the
non-equilibrium fluid dynamics problem. The second challenge involves solving the
equations numerically. Finally one would like 
to compare the results of non-equilibrium fluid dynamics to observables. This will
require an effort to solve the full system of the resulting equations
numerically.

The consequences of non-ideal fluid dynamics, both first-order (if applicable) 
and second-order were demonstrated here using a simplistic situation. 
We have seen that dissipative effects will be important in the early stages of the
collision dynamics. 
A more careful study of the effect
of the non-ideal fluid dynamics on the observables is therefore important. 
Conversely, measurements of the observables related to
thermodynamic quantities  would allow us to determine the importance and strength
of dissipative processes in heavy-ion collisions.

Here we have used only the truncated version of the transport equations and a
simple equation of state. 
A more realistic situation will require careful analysis of both the transport
coefficients and the equation of state which are employed in the full set of the
equations (including the terms omitted here). It is then that one may have a
better understanding of when to use either of these theories in the context of
relativistic heavy ion collisions. 
The study of non-ideal or non-equilibrium fluid dynamics will be important for
constructing hydro-molecular dynamic schemes \cite{azwi:bass}.  A resulting
hydro-molecular dynamic scheme can then be compared to the studies of multi-fluid
dynamics \cite{azwi:multifluid}.

\section*{Acknowledgements}
I would like to thank the organizers of the {\em 17th Winter Workshop on Nuclear
Dynamics} for inviting me to give this presentation. I would also like
to thank the participants for their comments and especially for the 
interesting and motivating discussions with S. Gavin. 
I would like to thank J.I. Kapusta for 
reading the manuscript and valuable comments. 
I would also like to thank D.H. Rischke, A. Dumitru, S.A. Bass 
for valuable comments.
This work was supported by the US Department of Energy grant
DE-FG02-87ER40382.

\vfill\eject
\end{document}